# A Constrained Path Quantum Monte Carlo Method for Fermion Ground States

Shiwei Zhang
*Center for Nonlinear Studies and Theoretical Division*
*Los Alamos National Laboratory, Los Alamos, NM 87545*

J. Carlson and J.E. Gubernatis
*Theoretical Division, Los Alamos National Laboratory, Los Alamos, NM 87545*

We propose a new quantum Monte Carlo algorithm to compute fermion ground-state properties. The ground state is projected from an initial wavefunction by a branching random walk in an over-complete basis space of Slater determinants. By constraining the determinants according to a trial wavefunction $|\Psi_T\rangle$, we remove the exponential decay of signal-to-noise ratio characteristic of the sign problem. The method is variational and is exact if $|\Psi_T\rangle$ is exact. We report results on the two-dimensional Hubbard model up to size $16 \times 16$, for various electron fillings and interaction strengths.

Quantum Monte Carlo (QMC) methods are often the only suitable tool for microscopic calculations of strongly-interacting many-body systems. They employ a stochastic means to sample the ground-state wavefunction or density matrix, from which ground-state or finite temperature properties can be computed. QMC methods have many different forms and are applied extensively to a wide range of problems in physics and chemistry. Applications to systems of fermions, however, are plagued by the so-called sign problem [1,2], which arises from the combination of the Pauli principle and the use of random sampling. The common signature of the sign problem is an exponentially vanishing signal-to-noise ratio as the system size or inverse temperature is increased. For years, this problem has significantly hindered simulations of many-fermion systems.

In this Letter, we propose a new QMC algorithm to study fermion ground-state properties. We cast the projection of the ground state as a branching random walk (RW) in an over-complete space of Slater determinants. This approach combines important advantages of two existing methods, the Green's function Monte Carlo (GFMC) [3,4] and the auxiliary-field quantum Monte Carlo (AFQMC) [5,6] methods. A constrained path (CP) approximation is then imposed on the determinants, which requires that the overlap with a trial wavefunction $|\Psi_T\rangle$ remain positive. The resulting method is free of *any* decay of the signal-to-noise ratio. It is variational and is exact if $|\Psi_T\rangle$ is exact. The CP approximation adopted here is based upon the positive projection approach of Fahy and Hamann [7], but can also be viewed as a generalization of the fixed-node (FN) [8–10] approximation in GFMC. Test applications of the algorithm to the two-dimensional Hubbard model yield, for the first time, very accurate results (energy and various correlation functions) for large systems.

Our general approach is independent of the form of the Hamiltonian, but we will use the familiar two-dimensional Hubbard model on a square ($N = L \times L$) lattice for illustrative purposes:

$$H = K + V = -t \sum_{\langle ij \rangle \sigma} (c_{i\sigma}^{\dagger} c_{j\sigma} + \text{h.c.}) + U \sum_{i} n_{i\uparrow} n_{i\downarrow}. \quad (1)$$

We use the imaginary-time propagator $G \equiv \exp(-\Delta\tau H)$ to project out the ground state $|\Psi_0\rangle$ from some initial state $|\Psi^{(0)}\rangle$. For small $\Delta\tau$,

$$G = e^{-\Delta\tau K/2} e^{-\Delta\tau V} e^{-\Delta\tau K/2} + \mathcal{O}(\Delta\tau^3). \quad (2)$$

As in AFQMC, the factor involving interactions can be mapped into a one-body form by a Hubbard-Stratanovic (HS) transformation: $\exp(-\Delta\tau V) = \sum_{\mathbf{x}} P(\mathbf{x}) B_V(\mathbf{x})$. The auxiliary field $\mathbf{x} = \{x_1, x_2, ...., x_N\}$ introduces a fluctuating potential at each lattice site, and $P(\mathbf{x})$ is the probability density function for $\mathbf{x}$. For simplicity we assume the discrete HS transformation [6,11], i.e., each $x_i$ is either 1 or $-1$ and $P(\mathbf{x})$ is a constant, $1/2^N$. The propagator is then decomposed into a single-particle operator form:

$$G = B_K \sum_{\mathbf{x}} P(\mathbf{x}) B_V(\mathbf{x}) B_K \equiv \sum_{\mathbf{x}} P(\mathbf{x}) B(\mathbf{x}). \quad (3)$$

Here $B_K = \exp(-\Delta\tau K/2)$ and any $B$ implies $B_\uparrow \otimes B_\downarrow$.

For an initial state $|\Psi^{(0)}\rangle$ not orthogonal to the ground state $|\Psi_0\rangle$, $G^n|\Psi^{(0)}\rangle$ leads to $|\Psi_0\rangle$ at large $n$. In AFQMC, $|\Psi^{(0)}\rangle$ is a single determinant and the path-integral $\langle\Psi^{(0)}|G^n|\Psi^{(0)}\rangle \equiv \sum_{\{\mathbf{x}\}} D(\mathbf{x}^{(n)}, \mathbf{x}^{(n-1)}, ...., \mathbf{x}^{(1)})$ is evaluated by the Monte Carlo (MC) method, in which the $n$ sets of auxiliary fields are sampled according to the overlap $D$. Our approach, instead, is to turn the propagation process into a RW in a space $\mathcal{D}$. A point in $\mathcal{D}$ is $|\phi\rangle = |\phi_\uparrow\rangle|\phi_\downarrow\rangle$, where each $|\phi_\sigma\rangle$ is obtained from the $N_\sigma$ single-particle orbitals on the $N$ lattice sites. Formally, our procedure resembles GFMC [3,4], but the RW in the latter is in configuration space.

The propagation is described by the iterative equation:

$$|\Psi^{(n+1)}\rangle = \sum_{\mathrm{x}} P(\mathrm{x}) B(\mathrm{x}) |\Psi^{(n)}\rangle. \qquad (4)$$

The antisymmetric wavefunction $|\Psi^{(n)}\rangle$ at any stage $n$ can be written as some linear combination of $|\phi\rangle$. In the MC process, $|\Psi^{(n)}\rangle$ is *sampled* by a finite ensemble of points $\{\phi_k^{(n)}\}$ in $\mathcal{D}$ called *random walkers*. A stochastic realization of (4) is in principle straightforward: A x is generated randomly for each walker, and then its Slater determinant $|\phi_k^{(n)}\rangle$ is propagated with $B(\mathrm{x})$. Repeating this procedure for all walkers generates a new population $\{\phi_k^{(n+1)}\}$ that samples $|\Psi^{(n+1)}\rangle$.

This naive sampling approach, however, is not practical because of large statistical fluctuations, hence an importance sampling scheme [3,4] is required. We define an importance function $O_T(\phi) \equiv \langle \Psi_T | \phi \rangle$ as the overlap of $|\phi\rangle$ with a trial wavefunction $|\Psi_T\rangle$. Typically, $|\Psi_T\rangle$ is in general a linear combination of Slater determinants. Equation (4) can now be transformed into an iterative relation involving $|\tilde{\Psi}\rangle$ rather than $|\Psi\rangle$, where $|\tilde{\Psi}(\phi)\rangle = O_T(\phi)|\Psi(\phi)\rangle$. To ensure that the underlying propagation remains unchanged, $P(\mathrm{x})$ for a path $|\phi'\rangle = B(\mathrm{x})|\phi\rangle$ must be modified:

$$\tilde{P}(\mathrm{x}) \propto P(\mathrm{x}) O_T(\phi')/O_T(\phi). \qquad (5)$$

The probability $\tilde{P}(\mathrm{x})$ for choosing x now depends on both the initial and final positions of the path, but the iteration can still be carried out as a RW: The population of walkers $\{\phi_k^{(n)}\}$ now represents $|\tilde{\Psi}^{(n)}\rangle$. For each walker, a x is sampled from $\tilde{P}(\mathrm{x})$, $|\phi_k^{(n)}\rangle$ is propagated by $B(\mathrm{x})$, and then $|\phi_k^{(n+1)}\rangle$ is assigned a weight $w(\phi_k^{(n)}) = \sum_{\mathrm{x}} \tilde{P}(\mathrm{x})$. The process is repeated for all walkers in the current population, and the resulting population represents $|\tilde{\Psi}^{(n+1)}\rangle$. We note that if an exact wavefunction is chosen as $|\Psi_T\rangle$, the normalization $w(\phi_k^{(n)}) \equiv$ const, i.e., walkers will have no branching.

In practice, $\tilde{P}(\mathrm{x})$ is difficult to sample directly. We circumvent this by sweeping through components of x one at a time: $B_V(\mathrm{x}) = \prod_{i=1}^{N} b_V(x_i)$. The importance sampling procedure is implemented for each $x_i$, where it is easy to sample $\tilde{p}(x_i) \propto O_T(b_V(x_i)\phi)/O_T(\phi)$ and to compute the normalization $\sum_{x_i=\pm 1} \tilde{p}(x_i)$. Walkers are stablized at suitable imaginary-time intervals by normalizing and re-orthogonalizing the single-particle orbitals, as in AFQMC [6]. Schemes to control population sizes and reduce weight fluctuations are similar to those used in GFMC [4].

The determinant RW approach has distinct advantages over AFQMC. It is a true ground-state method ($n\Delta\tau \to \infty$) that can be easily carried out with efficient sampling techniques. We have frequently used $n\Delta\tau$ as large as 500, and $\Delta\tau$ as small as 0.01. Also, at no extra cost, a better initial wavefunction $|\Psi^{(0)}\rangle$ in the form of multiple determinants can be employed to reduce equilibration time. Compared with GFMC, this approach automatically imposes antisymmetry by the use of determinants. It is plausible that the sign problem is reduced even without any approximation. Indeed, at half-filling *or* at $U = 0$, the approach is exact and completely free of the sign problem. In general, we expect that propagation with single-particle orbitals will be more effective than with isolated points in configuration space. Furthermore, as we discuss below, the calculation of off-diagonal expectations is much easier than in GFMC.

The most significant advantage of the determinant RW approach is perhaps the simple and practical implementation of the CP approximation, which prevents the exponential sign decay and provides a stable method. The sign problem occurs because of the symmetry between $|\Psi_0\rangle$ and $-|\Psi_0\rangle$ [7,12], which implies a symmetry between any pair of Slater determinants $|\phi\rangle$ and $-|\phi\rangle$. That is, the ground state resides in either half of $\mathcal{D}$, separated by a nodal plane $\mathcal{N}$ (defined by $\langle \Psi_0 | \phi \rangle = 0$) whose location cannot be specified *a priori*. In the limit of small $\Delta\tau$, imaginary-time paths in $\mathcal{D}$ are continuous. Therefore, if $|\Psi^{(0)}(\phi)\rangle$ is entirely on one side of $\mathcal{N}$, a determinant cannot cross the nodal plane $\mathcal{N}$ without touching it. Because of the $+/-$ symmetry, however, the total asymptotic contribution from any point on $\mathcal{N}$ is exactly zero in the path-integral. In other words, only paths confined in the half space affect $|\Psi_0\rangle$. In a MC simulation, no knowledge of crossing is available; paths are sampled individually according to the absolute values of their weights. Except in cases where special symmetry prohibits the crossing of $\mathcal{N}$ (such as the particle-hole symmetry for (1) at half-filling), paths that touch $\mathcal{N}$ grow exponentially in number compared to confined paths. Therefore, asymptotically the signal-to-noise ratio vanishes.

If the nodal plane $\mathcal{N}$ were known, we could simply constrain the RW in the correct half-space, and the MC procedure would become stable and yield exact results. Without precise knowledge of $\mathcal{N}$, we seek an approximate scheme with a trial nodal plane $\mathcal{N}_T$. From the path-integral analysis above and the apparent analogy with FN, it is natural [7] to require that walkers maintain a positive overlap with $|\Psi_T\rangle$. However, because of the different basis spaces used in GFMC and CPMC, the physical context and implications of the FN and CP approximations are different. In particular, the built-in antisymmetry and the over-completeness of $\mathcal{D}$ are expected to make CPMC behave rather differently.

In their AFQMC calculations, Fahy and Hamann [7] applied the condition of positive overlap with a $|\Psi_T\rangle$. The difficulty with their method is in the non-local nature of the propagation in AFQMC: any change of a HS field value affects *all* determinants that follow along the path.

In other words, updating of x at any imaginary-time $n$ requires evaluation of the overlaps at all future times, and the simultaneous updating of all fields is required to find acceptable paths. From the path-integral picture, it is clear that the amount of computation increases rapidly with system size or imaginary time.

With our RW realization of the propagation, updates and propagation are local in time. It is therefore straightforward to place the constraint $\langle\Psi_T|\phi\rangle > 0$ on each walker at each time $n$. The constraint defines $\mathcal{N}_T$ according to $|\Psi_T\rangle$. Any new walker that violates the constraint is given a zero weight and is thus discarded [13].

Once the RW has equilibrated, measurements can be made. For example, the ground-state energy $E_0 = \langle\Psi_0|H|\Psi_T\rangle/\langle\Psi_0|\Psi_T\rangle$, so the CP estimate of $E_0$ is

$$E_0^c = \sum_{n,k} w(\phi_k^{(n)})[\langle\phi_k^{(n)}|H|\Psi_T\rangle/\langle\phi_k^{(n)}|\Psi_T\rangle] \Big/ \sum_{n,k} w(\phi_k^{(n)}). \quad (6)$$

To compute the expectation of an operator $A$ that does not commute with $H$, $|\Psi_0\rangle$ must be used instead of $|\Psi_T\rangle$ in the estimator. This can be done using the principle of forward walking [14] as in GFMC: We consider any walker $\phi$ at time $n$ and its descendents $\{\phi'\}$ at $n+\eta$. The paths $\phi \to \phi'$ are distributed according to $O_T(\phi')/O_T(\phi)$. For each of these paths we can *back-propagate* $|\Psi_T\rangle$, i.e., operate the series of propagators on $|\Psi_T\rangle$ in reverse imaginary-time order. In contrast with forward walking in GFMC, where the computation of off-diagonal expectations can be difficult or even impossible, local and non-local operators are not distinguished here.

Within the restricted half-space defined by $\mathcal{N}_T$, the CP approach yields an eigenstate $|\Psi_0^c\rangle$ for $H$, i.e., $H|\Psi_0^c\rangle = E_0^c|\Psi_0^c\rangle$. Therefore, in this half-space $E_0^c = \langle\Psi_0^c|H|\Psi_T\rangle/\langle\Psi_0^c|\Psi_T\rangle \equiv \langle\Psi_0^c|H|\Psi_0^c\rangle/\langle\Psi_0^c|\Psi_0^c\rangle$. Since $|\Psi_0^c\rangle$ is an antisymmetric wavefunction, this variational estimate must have the upper bound property: $E_0^c \geq E_0$. We emphasize that, in contrast with lattice GFMC, the RW in space $\mathcal{D}$ becomes continuous as $\Delta\tau \to 0$, regardless of the discrete nature of the original system.

It was speculated in Ref. [7] that the quality of the approximation should improve as $|\Psi_T\rangle$ approaches $|\Psi_0\rangle$. We see that the physical meaning of $\mathcal{N}$ and the implication of the CP constraint are in fact rather clear. For example, the CP results become exact if $|\Psi_T\rangle = |\Psi_0\rangle$: Any $|\phi\rangle$ satisfying $\langle\Psi_0|\phi\rangle = 0$ can be expanded as a sum of the excited states of the system. Imaginary-time evolution of such a determinant will remain orthogonal to $|\Psi_0\rangle$. Therefore, its asymptotic contribution is zero. From (6), we also see that $E_0^c = E_0$ and the statistical error disappears.

To test the CPMC method, we calculated the energy and correlation functions for the two-dimensional Hubbard model. As in AFQMC, results are extrapolated to the $\Delta\tau = 0$ limit to remove the Trotter error from (2). For this purpose, calculations are carried out with at least three $\Delta\tau$ values (e.g., 0.07, 0.1, and 0.15 for $U = 4$, and smaller values for larger $U$). The trial wavefunction is either a free-particle one or the Hartree-Fock solution from an initial state in which the $\uparrow$ and $\downarrow$ electrons are placed on the two sublattices respectively. The calculations required CPU times ranging from a few minutes ($4 \times 4$) to about 50 hours ($16 \times 16$) on an IBM RS6000 590.

In Table I, our calculated ground-state energies and the variational values from $|\Psi_T\rangle$ are shown together with exact results, for various $4 \times 4$ systems. We note that $5 \uparrow 5 \downarrow$ corresponds to a closed shell case, while in $7 \uparrow 7 \downarrow$ the Fermi level falls in a degenerate set of free-particle states and the sign problem is quite severe in the AFQMC calculation. In fact, in this case it is rather difficult to obtain useful results with AFQMC, even at $U = 4$ [1].

A more stringent test of the algorithm is the calculation of correlation functions. Table II compares our results with available exact values. The structure factors $S_m$ and $S_d$ are the Fourier transforms of the real-space magnetic and charge density correlations, and $\rho(\mathbf{l}) = \langle\sum_{i,\sigma} c_{i+\mathbf{l},\sigma}^\dagger c_{i,\sigma}\rangle/N$. The MC data were obtained with the forward walking and back-propagation technique. The variational results are rather poor in most cases, implying that an extrapolation scheme [4] cannot be used. A poor $|\Psi_T\rangle$ also makes the forward walking less stable; the fact that our method works well is therefore even more significant.

In Fig. 1, we show the computed $E_0/N$ at $U = 4$ for various lattice sizes and electron filling $\langle n\rangle = (N_\uparrow + N_\downarrow)/N$. Also shown is available AFQMC data [17]. We see that our results, which are variational, are often indistinguishable from the AFQMC data, which should be exact. It is, however, the fundamentally different behaviors of the statistical errors that merit emphasis. In AFQMC the sign problem causes the variance to increase *exponentially* with $N$ and $n\Delta\tau$, while the CPMC method is stable and therefore exhibits power law scaling with $N$. For $10\times10$ systems, the CPMC error bars are 30-50 times smaller than those of AFQMC. For $L > 10$, the CPMC error bars are barely discernible, while AFQMC fails to yield meaningful results. Indeed, the only $12 \times 12$ data point by AFQMC not only has a large error bar, but also appears to lie *above* the CPMC value.

We also computed the momentum distribution, pair correlation functions, and structure factors for systems shown in Fig. 1. The forward-walking remains well-behaved, and the statistical errors very small. For smaller systems ($L = 6, 8, 10$), various comparisons with existing AFQMC data indicated that the agreement was similar to that in Table II. For instance, the structures of the computed $S_m(\mathbf{k})$ vs. $\mathbf{k}$ curves, including peak positions, were in exact agreement with AFQMC results [17].

Very recently, ten Haaf *et. al.* [10] constructed a standard GFMC FN approach for lattice fermions. It would be instructive to compare it with CPMC. The advantages that the determinant RW has over standard GFMC, namely built-in antisymmetry, over-completeness of the basis space, and easier calculation of off-diagonal expectation values, should still hold.

In conclusion, we have presented a variational, stable QMC approach for fermions. Very accurate results were obtained even with a simple $|\Psi_T\rangle$. Generalization to continuous HS fields is straightforward. We expect the method to be useful in a variety of applications, including continuum systems such as nuclei, atoms, molecules, etc. Algorithmic topics for further exploration include improvement of $|\Psi_T\rangle$ and development of released-node [9] and interacting-walker [12] analogs.

We are grateful to S.B. Fahy, M. Imada, and S. Sorella for helpful communications. S.Z. wishes to thank M.H. Kalos for encouragement, support, and stimulating discussions. Part of the calculations were performed on the Cornell MSC computing facilities. The work of J.E.G. and S.Z. was supported in part by the High Performance Computing and Communication program of the Department of Energy, and that of J.C. by the Department of Energy.

---

TABLE I. Comparison of variational, exact diagonalization, and our calculated (CPMC) ground-state energies per site for $4 \times 4$ systems. Exact values for $5 \uparrow 5 \downarrow$ systems are from Ref. [15], and those for $7 \uparrow 7 \downarrow$ are from Ref. [16].

| | $U$ | Variational | CPMC | Exact |
|---|---|---|---|---|
| $5 \uparrow 5 \downarrow$ | 4 | −1.1088 | −1.2238(6) | −1.2238 |
| $5 \uparrow 5 \downarrow$ | 8 | −0.7188 | −1.0925(7) | −1.0944 |
| $7 \uparrow 7 \downarrow$ | 4 | −0.8669 | −0.9831(6) | −0.9838 |
| $7 \uparrow 7 \downarrow$ | 8 | −0.592 | −0.728(3) | −0.742 |
| $7 \uparrow 7 \downarrow$ | 12 | −0.474 | −0.606(5) | −0.628 |

TABLE II. Comparison of selected values of correlation functions for $4 \times 4$ systems. $S_m$ and $S_d$ are the magnetic and density structure factors, respectively, and $\rho$ is the one-body density matrix. Exact diagonalization results are from Ref. [15].

| | $5 \uparrow 5 \downarrow, U = 4$ | | | $5 \uparrow 5 \downarrow, U = 8$ | | | $7 \uparrow 7 \downarrow, U = 4$ | | |
|---|---|---|---|---|---|---|---|---|---|
| | $\rho(2,1)$ | $S_m(\pi,\pi)$ | $S_d(\pi,\pi)$ | $\rho(2,1)$ | $S_m(\pi,\pi)$ | $S_d(\pi,\pi)$ | $\rho(2,1)$ | $S_m(\pi,\pi)$ | $S_d(\pi,\pi)$ |
| Variational | −0.125 | 0.624 | 0.624 | −0.125 | 0.625 | 0.625 | −0.103 | 4.4 | 0.517 |
| CPMC | −0.112(1) | 0.73(1) | 0.504(1) | −0.092(3) | 0.76(1) | 0.440(3) | −0.101(2) | 2.9(2) | 0.428(2) |
| Exact | −0.112 | 0.73 | 0.506 | −0.097 | 0.75 | 0.443 | −0.101 | 2.2 | 0.424 |

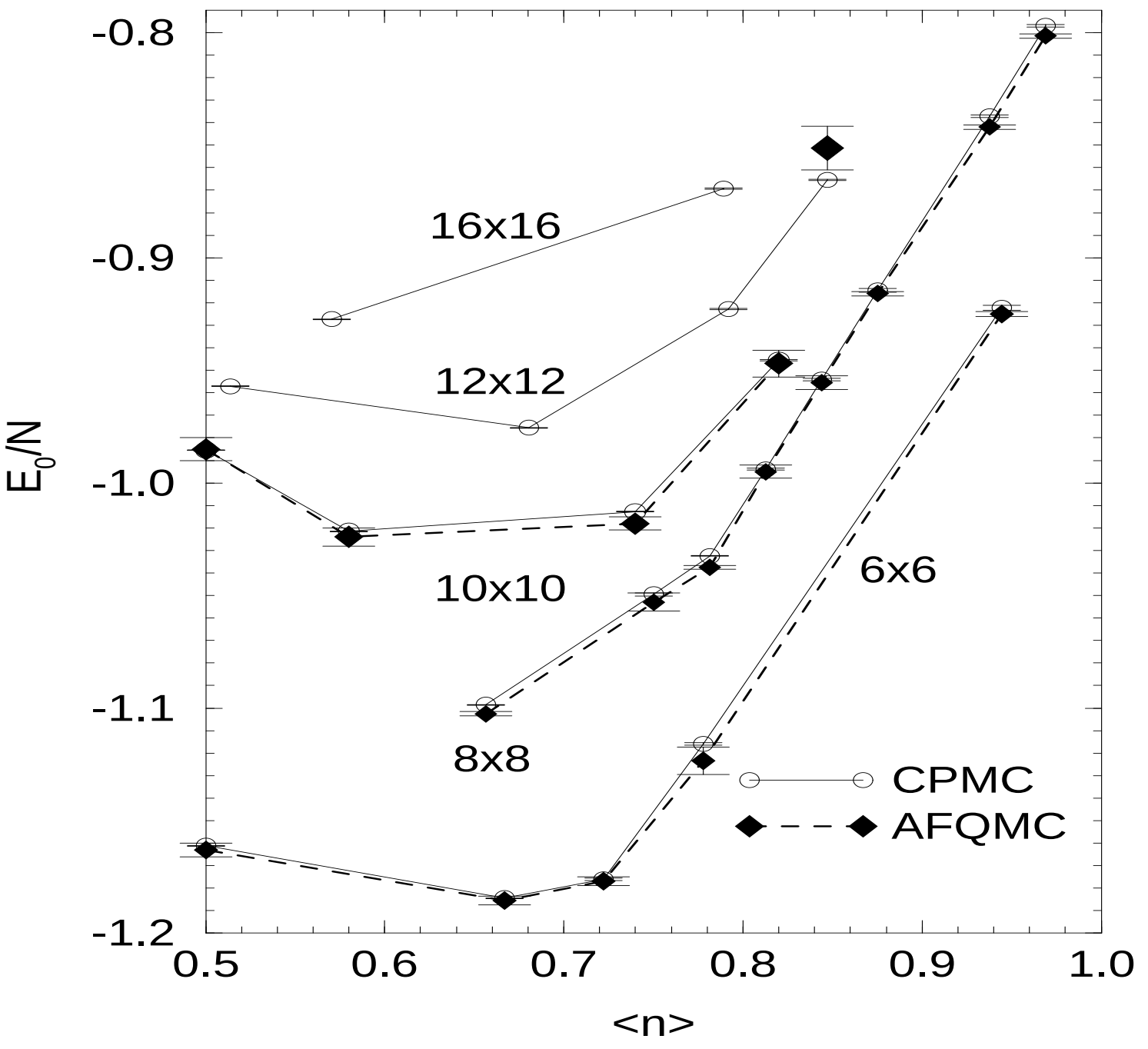

FIG. 1. Energies per site vs. electron fillings from CPMC, together with available AFQMC data. The lines are to aid the eye. Curves for $L = 8, 10, 12, 16$ are shifted up by 0.1, 0.15, 0.2, 0.25 respectively.